\documentclass[preprint]{iucr}              
\usepackage{graphicx}
\usepackage{rotating}
\usepackage{longtable}
     \paperprodcode{a000000}      
     \paperref{xx9999}            
     \papertype{SC}               

     \paperlang{english}          
     \journalcode{J}              
     \journalyr{2007}
     \journaliss{1}
     \journalvol{63}
     \journalfirstpage{000}
     \journallastpage{000}
     \journalreceived{0 XXXXXXX 0000}
     \journalaccepted{0 XXXXXXX 0000}
     \journalonline{0 XXXXXXX 0000}

\begin{document}                  

\title{2D anisotropic scattering pattern fitting using a novel Monte Carlo method: Initial results}
\shorttitle{Monte Carlo: now in 2D!}

\author[a]{Brian R.}{Pauw}{brian@stack.nl}
\author[b]{Masato}{Ohnuma}
\author[c]{Kenji}{Sakurai}
\author[d]{Enno A.}{Klop}

\aff[a]{International Center for Young Scientists, National Institute of Materials Science, 305-0047 Tsukuba, Japan}
\aff[b]{National Institute of Materials Science, 305-0047 Tsukuba, Japan}
\aff[c]{National Institute of Materials Science, 305-0047 Tsukuba, Japan}
\aff[d]{Teijin Aramid BV, Fibre Physics Group, PO 5153, 6802 ED Arnhem, The Netherlands.}

\shortauthor{Pauw, Ohnuma, Sakurai and Klop}

\maketitle

%

\begin{abstract}

Recently, a Monte Carlo method has been presented which allows for the form-free retrieval of size distributions from isotropic scattering patterns, complete with uncertainty estimates linked to the data quality. Here, we present an adaptation to this method allowing for the fitting of anisotropic 2D scattering patterns. The model consists of a finite number of non-interacting ellipsoids of revolution (but would work equally well for cylinders), polydisperse in both dimensions, and takes into account disorientation in the plane parallel to the detector plane. The method application results in three form-free distributions, two for the ellipsoid dimensions, and one for the orientation distribution. It is furthermore shown that a morphological restriction is needed to obtain a unique solution.

\end{abstract}


\section{Introduction}

Anisotropy in materials nanostructure is often induced upon fabrication or processing into industrially relevant shapes (e.g. wires, fibres, bars and sheets). For high-performance polymer fibres, for example, the oriented structure is effected by fibre spinning and further refined in tensioned heat-treatment procedures \cite{Northolt-1991}. Such an oriented structure has been shown to have a high correlation to the physical properties of the material \cite{Rao-2001a}. It is therefore of significant interest to extract as many structural parameters as feasible to achieve a good insight in the structure-property relationships for such anisotropic morphologies. Small-angle X-ray scattering can obtain statistically relevant bulk structural parameters for a plethora of such materials.

In small-angle scattering these anisotropic morphologies can give rise to anisotropic scattering patterns, which inherently contain an increased dimensionality of information. When such scattering patterns are sectioned, with each section treated as if it originated from an isotropic structure, invaluable cross-correlations are invariably lost. Fitting of a complete two-dimensional anisotropic scattering pattern is therefore of high interest. Prior work by others \cite{Fischer-2010,De-Geuser-2012} as well as this author \cite{Pauw-2010a} on describing such anisotropic scattering patterns based on classical\footnote{``Classical'' here implies the combination of a form factor with a structure factor and predefined size distribution, ofttimes fitted using a least-squares minimisation procedure commonly with up to 10 variables, as described in detail by \cite{Pedersen-1997}. Conversely, the ``generally applicable methods'' in this paper describe methods that retrieve (form-free) either the form factor, structure factor \emph{or} the size distribution when the remaining two are assumed.} descriptions of data have led to useful but complicated methodologies which tend to be rather inflexible and require tuning to the sample at hand. Development of a more flexible, generally applicable method is therefore rather important.

%

While for isotropic scattering patterns several such generally applicable methods have appeared over the years \cite{Glatter-1977,Magnani-1988,Krauthauser-1996}, for \emph{anisotropic} scattering patterns such form-free retrieval methods have been much more slow to develop. Of these, the inversion method by \citeasnoun{Stribeck-2001} in particular has been a significant milestone, capable of retrieving two-dimensional correlation function for anisotropic scattering patterns. Despite providing information in real space, however, correlation functions remain difficult to interpret. 

Shown here, therefore, is a modification to our recent MC method \cite{Pauw-2013} to allow for full fitting of anisotropic scattering patterns. This modification leads to the extraction of three form-free distributions, with two describing the anisotropic scatterers' dimensions and the third the orientation distribution. The first application results are discussed after a brief description of the necessary changes implemented to its isotropic forerunner. Parameter retrieval from a simulated scattering pattern shows that the original parameters are reasonably well retrieved. We think these are encouraging results and by presenting them we hope to foster discussions towards further development of the proposed MC method.

\section{Required adaptations to the MC method}\label{sc:imp}

The MC method proposed here is essentially identical to that previously described in \citeasnoun{Pauw-2013}. Differences can be summarised as: \textbf{1)} the implementation of a different elementary scatterer: an oriented ellipsoid of revolution with semi-axes $a$, $b=a$ and $c$ whose axis radii will be denoted by $R_a$ and $R_c$, respectively. \textbf{2)} The use of the detector angle $\psi$ (c.f. in \cite{Pauw-2010a}), and \textbf{3)} a single MC iteration uses \emph{three} randomly generated numbers that define $R_a$ and $R_c$, and the in-plane rotation $\psi_{\mathrm{ell}}$. For computational consistency (but without a change in function), the previously utilised ``compensation parameter $p_c$'' has been replaced by a ``radius-power factor'' $w$, where $w=\frac{6-p_c}{6}$. The changes in the affected MC procedural steps due to all aforementioned modifications are given below.

\subsection{Scattering of an ellipsoid}
The 2D scattering of an ellipsoid of revolution onto a detector pixel characterised by the scattering vector length $q$ and azimuthal angle $\psi$ can be calculated through modification of the Rayleigh function for a sphere (as used in previous work \cite{Pauw-2010a}). Here, $q$ is defined as $q=4\pi/\lambda \sin(\theta)$, where the radiation wavelength is denoted as $\lambda$, and $2\theta$ is the scattering angle, and the ellipsoids are considered oriented with their axes aligned perfectly parallel to the detector plane, and its main axis $c$ rotated from the detector $y$-axis by  $\psi_{\mathrm{ell}}$ :

\begin{equation}\label{eq:Iell}
F_\mathrm{ell}(q,\gamma,R_{a},R_{c})=3 \frac{\sin(qR_\mathrm{ell})-qR_\mathrm{ell} \cos(qR_\mathrm{ell})}{(qR_\mathrm{ell}^3)}
\end{equation}
where:
\begin{equation}\label{eq:Rell}
R_\mathrm{ell}(\gamma,R_{a},R_{c})=\sqrt{ R_a^2 \sin^2(\gamma) + R_c^2\cos^2(\gamma) } 
\end{equation}
and:
\begin{equation}\label{eq:psic}
\gamma=\psi-\psi_{\mathrm{ell}}
\end{equation}

The volume for an ellipsoid is given by:
\begin{equation}\label{eq:Vell}
V_\mathrm{ell}=\frac{4}{3}\pi R_a^2R_c
\end{equation}

\subsection*{Step 1: Preparation of the procedure}
The initial guess of the total scattering cross-section is calculated for a fixed number of ellipsoid contributions $n_s$ (commonly set to 500) whose radii $R_a$ and $R_c$ as well as the in-plane misalignment $\psi_{\mathrm{ell}}$ are randomly sampled from bounded uniform distributions. The radii sampling method is bounded by the radius range dictated by the measured q-range as detailed in the previous work \cite{Pauw-2013}, and the orientation distribution of ellipsoid axis $c$ is bounded by 45 degrees on either side of the fibre axis. It is important to note that the random sampling method is not quantised, so that the values can therefore assume any floating-point value within the size bounds. This initial guess is calculated using the general equation:

\begin{equation}\label{eq:main}
I_\mathrm{MC}(q)=B + A \sum\limits^{n_s}_{k=1} \mid F_\mathrm{ell,k}(q,\gamma_k,R_{e,k},R_{m,k}) \mid^2 \left(\frac{4}{3}\pi\right)^2 R_{e,k}^{2 w}R_{m,k}^w
\end{equation}

where the sum runs over all ellipsoids in the finite set. $B$ is a constant background term, and $A$ is a scaling factor, which is related to the volume fraction $\phi$ of the scatterers through:

\begin{equation}\label{eq:A}
A=\phi \Delta\rho^2 \sum\limits^{n_s}_{k=1} \frac {1}{\frac{4}{3}\pi R_{e,k}^{(2 w)} R_{m,k}^{(w)} }
\end{equation}
where $\Delta\rho$ is the scattering contrast. 
The volume fraction of the scatterers $\phi$ is defined as:
\begin{equation}\label{eq:vf}
\phi=\frac{V_\mathrm{scatt}}{V_\mathrm{irr}}
\end{equation}
where $V_\mathrm{irr}$ is the irradiated sample volume, and $V_\mathrm{scatt}$ the total scatterer volume in $V_\mathrm{irr}$.

$A$ and $B$ are tuned to the measured scattering through optimisation using a least-squares residual minimisation procedure, minimising the reduced chi-squared $\chi^2_r$ \cite{Pedersen-1997}, modified for 2D patterns:
\begin{equation}\label{eq:chisqr}
\chi^2_r=\frac{1}{N-M}\sum\limits^{N}_{i=1}\left[\frac{I_\mathrm{meas}(q,\psi)_i-I_\mathrm{MC}(q,\psi)_i}{\sigma(q,\psi)_i}\right]^2
\end{equation}
where $N$ denotes the number of data points, and $I_\mathrm{meas}(q,\psi)$ and $I_\mathrm{MC}(q,\psi)$ the measured and calculated model scattering cross-section at the position defined by $(q,\psi)$, respectively. $\sigma(q,\psi)_i$ is the the estimated error on each measured data point, whose estimation method is detailed in paragraph \ref{sc:dcor}. $M$ is the number of degrees of freedom in the fitting model, but is unfortunately ill-defined in an MC model and thus set equal to two for the scaling parameter $A$ and background contribution parameter $B$.

\subsection*{Step 2: Optimisation cycle}
Step two remains the same as for the isotropic MC variant, with the exception that every suggested replacement ellipsoid will be defined by three new sampled random variables. While it is possible to suggest changing only a single parameter of the original ellipsoid, this may have the potential to lead to biased optimisation and is therefore not done.

\subsection*{Step 3: Convergence and post-optimisation procedures}

After convergence has been reached, the partial volume fraction for each ellipsoid contribution $k$, $\phi_k$ is then calculated through reformulation of equation \ref{eq:A}:
\begin{equation}\label{eq:Vfk}
\phi_k=\frac{A\frac{4}{3}\pi R_{e,k}^{2 w} R_{m,k}^{w}}{\Delta\rho^2}
\end{equation}
where A is known through least-squares fitting from step 2. If this equation is calculated for all $n_s$ contributions, the volume fraction of the scatterers $\phi$ can be calculated as $\phi= \sum\limits^{n_s}_{k=1} \phi_k$.

\section{Experimental}

\subsection{Simulation details}
A scattering pattern was simulated using 1$\times 10^5$ ellipsoids over a grid of points whose $q$ and $\psi$ vectors are identical to that of the real measurement described below. The ellipsoids were assigned $R_a$ and $R_c$ values using a random sampling procedure over triangular volume-weighted distributions. The lower limit, mode and upper limit of the distribution for $R_a$ were 1, 1, and 20 nm, respectively. For $R_c$, these values were 1, 50 and 50, respectively. The orientation value for the ellipsoid axis $c$ was randomly sampled from a uniform volume-weighted distribution with lower and upper bounds of 80 and 100 degrees, respectively (in all scattering patterns, $\psi$ runs clockwise from zero at the 12 o'clock position).

The calculated intensity was subsequently scaled to a maximum intensity of 1$\times 10^6$ counts, quantised using a rounding procedure, and baseline noise of 0 or 1 count was added to all calculated pixel intensity values. The data was subsequently subjected to a 2x2 pixel binning procedure in order to get the data fitting times to within practical values (typically on the order of hours for 50 repetitions). Data below $q<0.12$\,nm$^{-1}$ is masked to resemble the real collected data, and data above $q>3$\,nm$^{-1}$ is also masked due to the unequal coverage of the angular region above this value (i.e. beyond $q\approx3$\,nm$^{-1}$ only the corner sections of the detector contain data, no longer a full circle).

\subsection{Data collection details}

SAXS measurements were performed on a laboratory instrument based around a rotating anode generator with a molybdenum target. Mo\,$k_{\alpha}$ emitted radiation was selected and focused using an Osmic confocal mirror. The incident beam was further collimated using three pinholes of 0.3, 0.2 and 0.45 mm in diameter, with the sample positioned immediately after the third pinhole. The distance from the optics to pinhole \#1, \#2 and \#3 is 0.115, 0.582 and 0.86\,m, respectively. Assuming that the second pinhole is the beam-defining pinhole, the transverse coherence length (and therefore the maximum size of precipitates contributing to the scattering signal) is on the order of 200\,nm \cite{Veen-2004}. Nominal beam diameter at the sample position is estimated to be approximately 200\,$\mu$m. The flightpath is a continuous vacuum from optics to detector.

For detection of the scattered radiation, a Dectris Pilatus 100k detector was used in combination with a translation stage. For every measurement, three images are collected and stitched together to form a single, almost square scattering image consisting of 487 by 497 pixels, each measuring 172\,$\mu$m by 172\,$\mu$m. The detector is placed 1.36\,m from the sample, its distance verified by ruler and measurement of a silver behenate standard, allowing for an angular coverage of $0.1$ to $4$\,nm$^{-1}$ in $q$. As the beamstop is mounted on thin polymer wires, there is no significant shadow from a beamstop holder, and thus no gap in angular coverage. 

The sample measurement was performed for 14 hours per frame, with three frames making up the scattering image. The background (no sample) was measured for 6$\frac{2}{3}$ hours per frame. The transmission factor was measured by removing the beamstop and measuring the attenuated direct beam ($I_0$) and the attenuated beam with the sample in position ($I_1$), the difference of which $I_1/I_0$ results in a transmission factor of 0.93(5). 

\subsection{Sample details}

A commercially available sample of PPTA fibre (Twaron 1000) was obtained from Teijin Aramid B.V., with a gravimetric density of approximately 1450(20)\,kg/m$^3$ and an elemental composition of C$_7$H$_5$NO. This fibre is wound on a small rectangular frame of 13 by 18 mm used to standardise WAXS and SAXS measurements, as previously reported \cite{Pauw-2010a}. Five such frames were packed together in order to maximise the scattering signal and average over a larger quantity of material, leading to a total thickness of aramid along the beam of 0.671\,mm. As the fibres tend to carry moisture from the air, they were subjected to a $>$24\,h drying process in a vacuum desiccator prior to the measurement. 

\subsection{Data correction}\label{sc:dcor}

In order to achieve as high as possible relative data accuracy, the collected data is corrected for natural background radiation, transmission, sample thickness, measurement time, primary beam flux, parasitic background, polarisation, detector solid angle coverage (angular dilation) and sample self-absorption using in-house developed data reduction software written in the Python language. Deadtime correction is unnecessary at the count rates encountered in this study \cite{Kraft-2009}, and dead-pixel and flatfield corrections are performed by the detector acquisition software prior to ingestion by the data reduction software \cite{Eikenberry-2003}. 

The collected intensity is binned in blocks of 2 by 2 pixels to improve calculation speed. Data below $q<0.12$\,nm$^{-1}$ and above $q>3$\,nm$^{-1}$ is masked (\textit{ibid.}). 

The data is furthermore scaled to absolute units using a calibrated glassy carbon standard \cite{Zhang-2010a}. Statistical uncertainty on the datapoints is set to the maximum of either: 1) The propagated Poisson counting statistics, 2) 1\% of the intensity in the histogram bin (the realistically minimum achievable uncertainty given the aforementioned corrections \cite{Hura-2000}) or 3) the standard deviation of the intensity values of the pixels in the bin. As the parasitic scattering is minimal due to complete beam path evacuation, improved statistics have been obtained by skewing the time division between sample and background measurement to favour sample measurement time \cite{Pauw-B2012}. Furthermore, the minimal parasitic scattering improves the data accuracy by its reduced sensitivity to inaccuracies in the transmission factor determinations during background subtraction.

\subsection{MC method settings}

The number of contributions $n_s$ has been set to 500, and w has been set to $0.5$. The scattering contrast is assumed to be the contrast between the aforementioned aramid and vacuum, resulting in a scattering contrast $\delta\rho^2$ of 1.64$\times10^{30}$\,m$^{-4}$. 
The radius bounds for $R_a$ and $R_c$ for the random number generator are set to $0.7\leq R\leq70$\,nm, the upper limit is set high as it is related to the actual beam width rather than the minimum measured $q$\footnote{Imagine, for example, a high aspect ratio prolate ellipsoidal scatterer. Its scattering pattern will resemble a streak in the detector plane, the width of which is (at minimum) the beam width. Thus, the maximum R is dictated by the beam spot width rather than the minimum measured $q$.}. 50 independent converged solutions are calculated for each fit to obtain uncertainty estimates on the resulting distributions.

\section{Results and discussion}\label{sc:rnd}

\begin{figure}
   \centering
   \includegraphics[angle=0, width=0.99\textwidth]{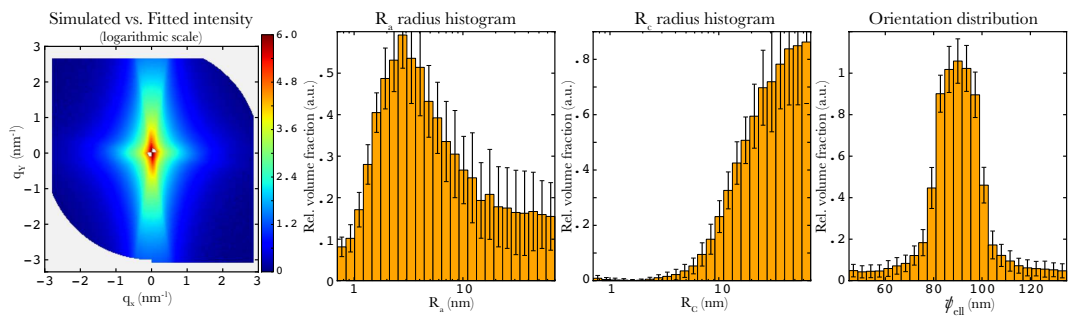} 
   \caption{Leftmost: 2D Monte Carlo fit (quadrants with rounded corners) to simulated data, with the (simulated) ``fibre'' axis horizontal, and intensity shown on logarithmic scale. Middle left: MC retrieved size histogram of the $a$-axis radius distribution originally simulated with lower limit, mode and upper limit of 1, 1 and 20 nm, respectively. Middle right: MC retrieved size histogram of the $c$-axis radius distribution, originally simulated with lower limit, mode and upper limit of 1, 50 and 50 nm, respectively. Rightmost: MC retrieved $c$-axis orientation distribution retrieval from simulated data, originally simulated using a uniform distribution with lower limit of 80 and upper limit of 100 degrees (90 degrees is horizontal). Distributions shown as volume-weighted distributions.} 
    \label{fg:sim1}
\end{figure}

The simulated data retrieval shows very good agreement of the calculated intensity with the simulated intensity as shown in Figure \ref{fg:sim1}, fitting on average to within the data uncertainty (i.e. a $\chi^2_r\leq 1$). In addition, the distributions are reasonably well retrieved, though some of the sharper details of the distributions used for the simulation have been lost. 
This can most likely be ascribed to the interplay of the somewhat similar effects of aspect ratio and orientation distribution for ellipsoidal scatterers, though a similar loss of sharp details has also been observed in the isotropic retrieval studies, especially for broad distributions (c.f. SI of \cite{Pauw-2013}). As the intensity is well described, however, the result is considered to be a reasonable estimate of the actual structural parameters.
 
\begin{figure}
   \centering
   \includegraphics[angle=0, width=0.99\textwidth]{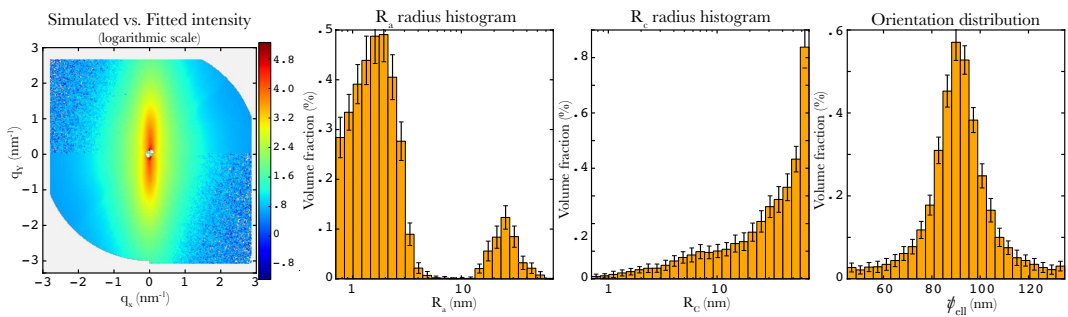} 
   \caption{Leftmost: Measured data versus MC result (quadrants with rounded corners) of Twaron 1000, with the fibre axis horizontal and the intensity on a logarithmic scale in absolute units of reciprocal meters. Middle left: $a$-axis radius distribution. Middle right: $c$-axis radius distribution. Rightmost: $c$-axis alignment (fibre axis horizontally at 90 degrees). Error bars indicate $\pm$ 1 standard deviation from the mean calculated from the result of 50 MC optimisations. Distributions shown as volume-weighted distributions.} 
    \label{fg:result}
\end{figure}

The Monte Carlo procedure as applied to analyse the scattering pattern effected by the pore structure of a PPTA fibre is shown in Figure \ref{fg:result}. The MC method can fit the data on average to within the data uncertainties (i.e. with a $\chi^2_r<1$). The retrieved size distribution shows few features in the $c$-axis size distribution, which may indicate that the maximum of that distribution lie well beyond the limits of the instrument. The $a$-axis distribution shows two clear features, one distribution at around 2 nm, and a second one at 20 nm. This result would indicate that there is a bimodal porosity in the Twaron 1000 fibre. Note that the real fibre-symmetric three-dimensional void orientation function is approximated here using a two-dimensional $c$-axis orientation distribution, similar to the treatment described in \citeasnoun{Pauw-2010a}.

\begin{figure}
   \centering
   \includegraphics[angle=0, width=0.95\textwidth]{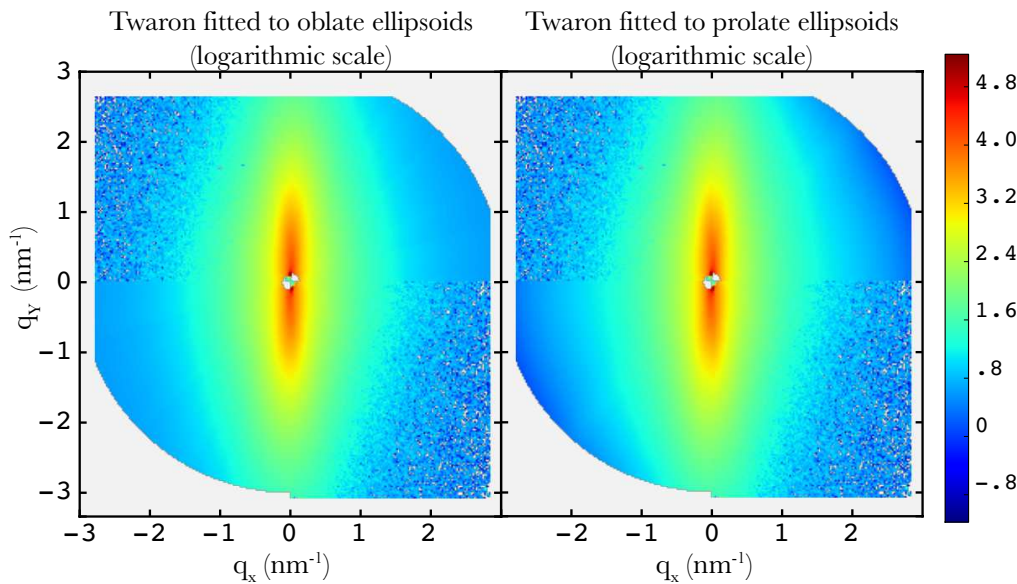} 
   \caption{Twaron 1000 measurements compared to the MC results (quadrants with rounded corners) for oblate ellipsoids (Left, with the $c$-axis within 45 degrees of the vertical axis) and prolate ellipsoids (Right, with the $c$-axis within 45 degrees of the horizontal axis). Intensity shown on logarithmic scale.} 
    \label{fg:result2}
\end{figure}

In the prior evaluation of the isotropic data retrieval using the MC method, it was evident that shape and size distribution cannot both be uniquely identified in a scattering pattern from polydisperse systems. A similar limitation exists for anisotropic scattering patterns, despite the availability of additional dimensionality in the data. To illustrate this, two solutions were calculated for the Twaron 1000 fibre, one consisting of oblate ellipsoids, the second consisting of prolate ellipsoids\footnote{The deciding factor for which shape was preferentially obtained is the bounds of the $c$-axis alignment. If the $c$-axes are allowed to vary within $\pm$45$^\circ$ from the direction perpendicular to the fibre axis (and parallel to the detector plane), oblate ellipsoids are preferentially obtained. If the $c$-axis axes are allowed to vary within  $\pm$45$^\circ$ \emph{parallel} to the fibre axis, however, prolate ellipsoids are obtained.}. Both solutions converge on average to within the uncertainty of the data, and are shown in Figure \ref{fg:result2}. 

These results stress the need for supplementary information when finding a solution to a small-angle scattering problem. In this case knowledge on the general scatterer shape (oblate or prolate, ellipsoids or cylinders) is needed from either microscopic investigations or considerations on the production process. Given the production process of the fibres, the existence of prolate ellipsoidal voids in the fibres is the obvious candidate.

The thus retrieved data particularly lend itself to further scrutiny, as it contains the interrelations of both dimensions (and thus the aspect ratio) alongside the orientational information. Depending on the requirements of the researcher, therefore, the ellipsoid parameters resulting from the MC method can be plotted against one another in a variety of ways. Amongst others, this allows for the study of the correlation between the aspect ratio and particle volume, or radius distribution and orientation angle. More simple parameters can also be obtained, such as the radius of gyration, the modes of the orientation distribution, and the total volume fraction of the phases. As such, we believe this is a very powerful and promising method of 2D data analysis.

\section{Upcoming improvements}

In contrast to the previously published isotropic MC method, this 2D variant can shamelessly be called ``very slow''. A single MC optimisation can take up to several hours on a single core of a 1.8 GHz intel i7 (as found in the 2011 Macbook Air), a similar timescale as the laboratory measurement itself. This appears mostly due to the size of the dataset (on the order of a quarter million datapoints, reduced to about 60000 upon 2x2 binning) and the need to repeatedly evaluate the ellipsoid scattering function as well as the chi-squared function with these points. One upcoming improvement therefore is the implementation of faster calculation functions for the ellipsoid scattering pattern, similar to those recently suggested by \citeasnoun{Littrell-2012}. Furthermore, multicore processing will be implemented to leverage more computing power. 


Secondly, the issues arising from the interplay between the similar effects of aspect ratio and orientation distribution may be further suppressed when cylindrical shapes are used instead. Cylindrical shapes have scattering patterns which are more direction-dependent than their ellipsoidal counterparts, and contributions arising from their aspect ratio should therefore be better separable from the orientation distribution effects. Alternatively, a Bayesian approach could be considered in which prior information from other techniques is used to arrive at a inter-technically consistent solution.

\section{Conclusions}

This report shows the applicability of MC methods to anisotropic scattering patterns, which can be used to extract all available information from a scattering pattern. A wide range of morphological questions can be answered through careful interpretation of the MC result. While the initial results are encouraging, improvements abound, in particular on improving the speed and accuracy of the method.

The MC code as well as the measured datasets are available for inspection, improvements and application under a Creative-Commons Attribution Share-alike license and the latest copy will be freely supplied by the author upon request.

\referencelist[/Users/brian/Documents/bibliography] 

\end{document}